# Precision Projector Laboratory: Detector Characterization with an Astronomical Emulation Testbed

Charles Shapiro[1], Roger Smith[2], Eric Huff[1], Andrés A. Plazas[1], Jason Rhodes[1,2], Jason Fucik[2], Tim Goodsall[1], Richard Massey[3], Barnaby Rowe[1,2,4], Suresh Seshadri[1]

[1] Jet Propulsion Laboratory, California Institute of Technology, Pasadena, CA, 91109
[2] California Institute of Technology, Pasadena, CA, 91125
[3] Durham University, Durham, DH1 3LE, United Kingdom
[4] University College London, London, WC1E 6BT, United Kingdom

**Abstract.** As astronomical observations from space benefit from improved sensitivity, the effectiveness of scientific programs is becoming limited by systematics that often originate in poorly understood image sensor behavior. Traditional, bottom-up detector characterization methods provide one way to model underlying detector physics, and generate ever more faithful numerical simulations, but this approach is vulnerable to preconceptions and over-simplification. The alternative top-down approach is laboratory emulation, which enables observation, calibration, and analysis scenarios to be tested without relying on a complete understanding of the underlying physics. This complements detector characterization and simulation efforts by testing their validity. We describe a laboratory facility and experimental testbed that supports the emulation of a wide range of mission concepts such as gravitational weak lensing measurements by WFIRST and high precision spectrophotometry of transiting exoplanets by JWST. An Offner relay projects readily customizable "scenes" (e.g. stars, galaxies, spectra) with very low optical aberration over the full area of a typical optical or near infrared image sensor. f/8 and slower focal ratios may be selected, spanning those of most proposed space missions and approximating the point spread function (PSF) size of seeing limited ground based surveys. Diffraction limited PSFs are projected over a wide field of view and wavelength range to deliver highly predictable image patterns down to sub-pixel scales with stable intensity and fine motion control. The testbed enables realistic validation of detector performance on science-like images, which aids mission design and survey strategy, as well as targeted investigations of various detector effects.

**Keywords:** Instrumentation, Calibration, Astronomy

## 1 Introduction

### 1.1 Detector Systematics In Astronomy

Many areas of astronomy have become increasingly dependent on measurement precision. This trend is driven in part by larger telescope apertures with larger, more sensitive focal plane arrays, which rapidly collect astronomical data, allowing random errors in scientific measurements to be averaged down to levels comparable to the instrument systematics. Calibrations that attempt to reduce residual systematic errors are aided by adaptive optics or by the stability of observations from space. Space-based instruments benefit from excellent thermal stability, uninterrupted by the diurnal cycle or weather, as well as a lack of atmospheric emission, extinction, scattering, or point spread function (PSF) degradation. The infrared background can also be greatly reduced by absence of atmospheric emission and lower telescope temperatures.

In this context it is increasingly common for space missions or large ground-based survey telescopes to place stringent requirements on post-calibration systematics. These require a detailed understanding of second order detector effects (reviewed briefly below) well beyond the standard flat fielding, dark subtraction, and linearity corrections. The impacts (and sometimes the mechanisms) of these second order effects are not fully understood, and standardized calibration procedures have not been widely implemented, since we are only recently recognizing their importance in high precision regimes. The propagation of the resulting errors to science metrics can have a complicated dependence on the image scale (for pixel-scale effects), observing sequence (for image memory effects), and calibration and analysis methods (e.g. combination of undersampled images). Moreover, these effects not only contribute their own measurement biases, but they can also bias, couple, and thus complicate the interpretation of the standard calibrations.

In charge-coupled devices (CCDs) for example, images are degraded during readout by charge transfer inefficiency. The amount of charge trapped and later released depends on pixel position within the CCD, signal and sky levels, and radiation damage history. This effect is well known but requires careful modeling to satisfy e.g. shape measurement requirements for weak gravitational lensing surveys [1]. Furthermore, charge collection is guided by not-entirely-vertical electric fields, creating pixel size variations that depend on absorption depth and thus wavelength. These fields may be sourced by obvious features such as bloom stops or the array edges, or more subtly by impurity gradients in the silicon substrate (the tree-ring effect in Dark Energy Survey CCDs) [2-4]. Pixel boundary locations also shift in response to clock voltages and signal gradients, giving rise to a "brighter-fatter" effect (fluence dependent PSF) discussed below [5,6].

In infrared Complementary Metal-Oxide-Semiconductor (CMOS) detectors, such as the ubiquitous Teledyne Hawaii-xRG (HxRG) family of HgCdTe devices, one must contend with interpixel-capacitance (IPC) which may be anisotropic [7] and nonlinear [8], image persistence [9,10], non-linearity in the charge to voltage conversion, and flux dependent sensitivity ("reciprocity failure") [11,12,13]. As we will discuss, pixel boundary shifts similar to the CCD effects mentioned above also seem to occur in HxRG detectors [14]. A significant further complication for infrared detectors is that all these properties vary from pixel to pixel, along with electrical gain, offset, dark current, quantum efficiency, and noise. Many properties are also sensitive to temperature and operating voltages, while persistence and reciprocity failure are dependent on exposure duration, cadence, and illumination history [13]. Furthermore, both CCDs and infrared detectors may exhibit variations in pixel size [15,16], position [17], and intra-pixel sensitivity [18,19], which cannot be calibrated by conventional flat fielding.

Many of these effects bias photometric, astrometric, spectroscopic, and shape measurements at the 1% level or less, so may not be routinely corrected by observatory-level calibration, but rather addressed as needed by science teams who anticipate an unacceptably large impact on their specific observables. While numerical simulations can be used to test sensitivity to a known effect, they are vulnerable to conceptual and implementation errors stemming from limitations in our understanding of the detector physics. For instance, an incorrect assumption about the order in which an effect occurs in the signal chain will lead to a flawed calibration procedure, which may nevertheless work well in the simulations!

There is an important role for end-to-end laboratory *emulation* of observations in which realistic scenes, or at least grids of point sources with the intended PSF, are projected onto representative image sensors. Emulation can evaluate the impact of various detector effects on science metrics with proposed calibrations applied, using realistic observing cadences, intensities, and analysis methods. Such tests can identify unexpected effects (including new effects not listed above) soon enough to affect sensor procurement and mission planning. Emulation can demonstrate the relative merits of different calibration strategies. By contrast, surprises in detector operation that are discovered in the science data risk costly delays and compromises to the science reach. Even when detector effects can be mitigated through data analysis methods, developing such methods can take years.

Whereas science requirements flow down to *post-calibration* detector performance, specifications provided to vendors must be expressed in terms of *raw* data obtained using the vendor's standard test methods. Otherwise, the vendor is unable to predict yields, and thus costs, without a *prior* test program tailored to each science case. The manufacturer's detector characterization procedures typically use flat illumination and dark frames and seldom probe phenomena that depend on the scene contrast or optical PSF. Therefore the responsibility of validating devices at the level needed for high precision measurements usually falls to those building or using the instruments. The Large Synoptic Survey Telescope (LSST) is an example of a ground-based survey that will achieve high statistical precision and whose original proponent, Tyson [20], has implemented experiments to emulate observations with the same focal ratio, passband and similar PSF to examine systematic galaxy shape measurement errors caused by the sensors.

**1.2  Preview**

This paper describes a detector characterization facility -- dubbed the Precision Projector Laboratory (PPL) – intended to address detector-related risks to ongoing or proposed space missions through emulation experiments. The PPL testbed (henceforth "the projector") is a one-to-one re-imager that focuses customized astronomical "scenes" (e.g. stars, galaxies, spectra) onto detectors. It operates from 0.3-2 µm and at focal ratios of f/8 or slower with very low aberrations over a 40mm square field of view, an area large enough to cover typical large-format image sensors. The ability to focus >$10^4$ images per scene provides a large multiplex advantage over single-spot scanners, enabling rapid mapping of photometric, astrometric, and PSF shape variations over an entire detector as opposed to selected pixels. In section 2, we discuss the design and performance of the projector testbed. In section 3, we describe examples of emulation experiments that demonstrate the capabilities of the projector and its value in detector characterization and mission planning.

PPL was initially created to demonstrate the feasibility of galaxy shape measurements in the near infrared (NIR) for space-based weak gravitational lensing surveys such as WFIRST [21]. However, a wide range of instruments and stimuli can be straightforwardly emulated by adjusting the scene, wavelength, focal ratio, and image motion. PPL has investigated the precision achievable for spectrophotometry of transiting exoplanets with the infrared detectors on the James Webb Space Telescope (JWST) [22,23] and the proposed Fast Infrared Exoplanet Spectroscopy Survey Explorer (FINESSE). It has also assessed the impact of sub-pixel response variations on photometry for the Euclid mission. PPL has proven useful for testing its own image stabilization system and the optical quality of a full field infrared re-imaging camera for the Keck telescope's adaptive optics system [24]. It also demonstrated the efficacy of an optimal recombination algorithm for undersampled images using a NIR detector sampling spatially at 1/2 of the Nyquist rate [25,26]. PPL was recently used to test the Wafer-Scale imager for Prime (WaSP) guider for the 200" Hale Telescope at Mt. Palomar, and it will soon characterize CCDs for the SuperBIT balloon-borne telescope [27].

We use representative detectors, projecting scenes resembling intended observations, with an emphasis on predictability of image properties such as PSF size, shape, intensity, and position. Realistic exposure cadences and pointing dither patterns can also be emulated. We then extract relevant parameters in analyses that are relevant to the science requirements of a given project. Results can be compared to requirements (or expectations) with varying degrees of calibration complexity. Candidate detectors with differing levels of performance can be compared to assess how the detector properties propagate through the whole data acquisition and analysis process. A program of emulation, calibration, analysis, and interpretation is successfully integrated thanks to a multi-disciplinary collaboration of astronomers, detector experts, and optical engineers at Jet Propulsion Laboratory and Caltech Optical Observatories. External collaborators are encouraged to visit Pasadena to operate the projector themselves and iterate on the execution and interpretation of the experiments with the PPL team.

## 2 The "Projector" Testbed

### 2.1 Opto-mechanical Layout

The projector needs to form an image of input masks onto the detector using a PSF relevant to the instrument to be emulated, and ideally a factor of two larger or smaller, so that dependencies on PSF scale can be probed. A magnification near unity is ideal since higher magnification tends to place excessive requirements on mask resolution and demagnification leads to larger mask sizes that drive up the size of the required optics. The 1:1 ratio chosen is a good match to the mask resolution and size commonly available from semiconductor mask makers. Masks are described further in 2.3.

The optical design, an Offner relay, is illustrated in Fig. 1 and Fig. 2. A W-shaped light path makes two reflections off a concave spherical mirror (M1/M3) with 1.5 m radius of curvature and 470 mm aperture diameter, with an intervening reflection from a 108 mm diameter convex mirror (M2) with 0.75m radius of curvature. The two curved mirrors were fabricated from Schott Zerodur substrates and polished to better than λ/100 (at 633nm) RMS wavefront accuracy. The Offner relay is correctly aligned when the centers of curvature of the two spherical mirrors are coincident, and it derives excellent optical performance over a wide area from the cancellation of aberrations due to symmetry of the input and output paths. When the misalignment of the centers of curvature is less than 25µm, the Strehl ratio is greater than 90% and PSF ellipticity is less than 0.1% rms over a 20mm x 20mm field at f/8 and 40mm x 40 mm at >f/11. The necessary alignment was achieved through initial placement of components to better than 25µm accuracy using a coordinate measuring machine (i.e. Faro Arm) followed by adjustment of M2 position and tilt using an alignment telescope looking through the light path at a retro-reflecting "tooling ball" located at the output focus. We added a fold mirror that reflects the beam upward through a turntable where we mount the detector being tested; this enables 360 degree rotation of the detector about the optical axis with constant gravity vector.

The focal length is fixed at 750 mm, and focal ratios from f/8 and slower are selectable by changing pupil stops, enabling a range likely to be used in space missions of interest. Sub-pixel spots can be produced for most wavelengths and pixel sizes of interest. Pupil stops can be made circular or mildly elliptical. Although ground based survey telescopes (e.g. Large Synoptic Survey Telescope, Zwicky Transient Factory) often operate at focal ratios faster than f/8, atmospheric seeing enlarges their PSFs. Seeing limited images can be emulated by some combination of diffraction (smaller pupil stop) and precise image motions during the exposure, produced by a tip-tilt mirror with piezo-electric actuators. Both the pupil stop and detector can rotate about the optical axis, allowing partitioning of PSF contributions between the detector and optics.

The optical PSF is well-approximated by an Airy disk with full-width-half-maximum given by $\text{FWHM} = 1.03\,\lambda\,F$, where $\lambda$ is the illumination wavelength and $F$ is the f-number. PSF measurements, when we include the detector response, are slightly blurred by lateral charge diffusion. PSF size is controlled via the f-number and wavelength to obtain various degrees of sampling from a detector with fixed pixel pitch. The sampling factor $Q$ for a given detector with pixel pitch $p$ is given by $Q = \lambda\,F/p$, where Q<2 (or roughly, FWHM<2 pixels) denotes an undersampled image, i.e. the pixel pitch does not achieve the Nyquist rate set by the optical band limit. When detector effects are insensitive to the wavelength and focal ratio of an instrument, matching $Q$ creates a PSF at the correct scale to assess the impact of pixel-scale detector effects on photometry, spectroscopy, and imaging.

A disadvantage of the Offner optical design is that its pupil stop is the M2 surface so that any aperture stop mask must be placed within 100 µm of the M2 surface. Furthermore, the (nominally black) pupil mask scatters light toward the primary, which reduces image contrast. The Offner relay remained the best design choice as it delivers very low aberrations over a wide object area and range of wavelengths. The magnitude of the aberrations remains approximately constant along one direction of the image. Refractive designs such as a Cooke Triplet for a collimator and camera were also considered.

Disadvantages of refractive designs are the limited wavelength range, chromatic aberrations, optical surfaces producing ghost images, and the radial distribution of aberrations.

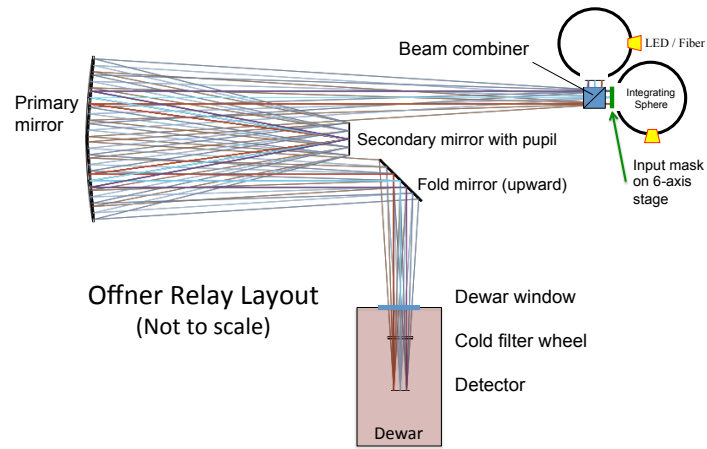

Fig. 1. Optical layout of the projector, a 1:1 magnification Offner relay with 750 mm focal length.

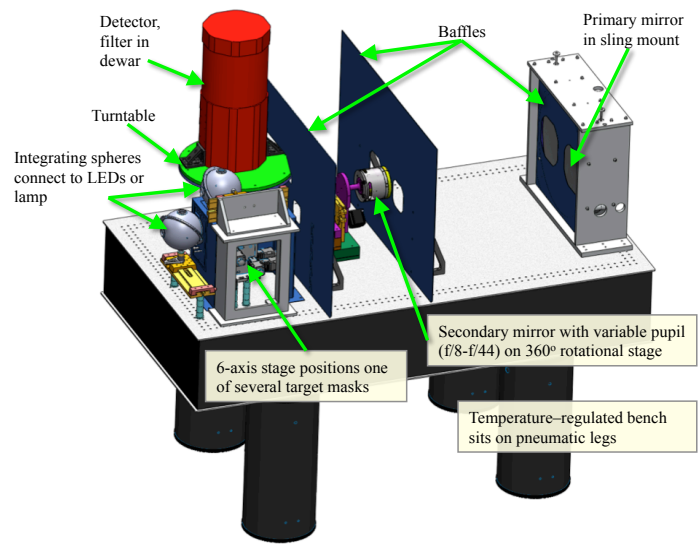

Fig. 2. Model of the projector's optical system. The system sits in an enclosure that shields background light and supports the light source and other equipment. Test devices are mounted looking down through the turntable.

## 2.2 Illumination

High power illumination is needed to enable short exposures and thus fast data acquisition of thousands of images. Bright targets can also be used to overcome NIR backgrounds and avoid image motion that would become problematic at longer timescales. Little light is passed by the 3 µm pinhole masks employed in many tests, (the light is spread over the much larger PSF), and throughput decreases at slow focal ratios since these are achieved by stopping down the pupil (flux at f/44 is 3% of flux at f/8). Flux can be attenuated in long exposures by turning down illumination power sources or by blocking/filtering the input to the integrating spheres. Most aspects of the illumination configuration are scriptable via RS232 connections. A flowchart of the hardware is shown in Fig. 3 and described in detail below.

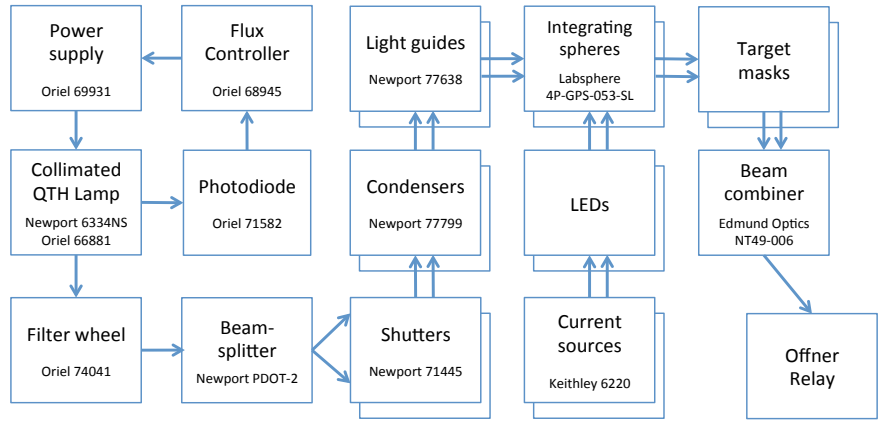

Fig. 3. Flowchart of the projector illumination system with part numbers. Arrows denote flow of light (optics) or control (electronics). The beam combiner may be swapped for a simple flip mirror.

For broad wavelength coverage extending from near UV to NIR (0.3-2 µm), two 5.3" internal diameter integrating-spheres with spectralon coatings are fed via liquid light guides from a high power (max 250W) quartz-tungsten-halogen lamp located outside the projector enclosure. The collimated lamp beam is evenly divided by a beamsplitter before focusing onto the light guides. Electronic shutters select one or both spheres, allowing us to combine scenes or quickly switch between a science target and flat calibration source. A filter wheel before the beamsplitter selects UV, visible, or NIR bandpass filters. For NIR detectors, a matching cold filter is needed to block blackbody radiation emitted by the room temperature optics; thus, a change in illumination wavelength requires a corresponding change of the filter wheel within the dewar. Alternatively, high power LEDs mounted directly to both integrating spheres provide 880 nm light (20 nm passband). This wavelength was chosen since it was the longest for which high power LEDs were available, and it lies within the passband of both HgCdTe sensors and CCDs. High power modules at 940 nm have since become available for purchase.

A silicon photodiode directly measures the lamp intensity by observing the filament through a small opening in the lamp housing. This measurement feeds into a flux controller, which in conjunction with the programmable power supply, achieves 0.1 % rms intensity stability over one day according to the manufacturer. We measure a peak to peak variation of 0.04% over 3 hours (after powering on and settling) sampling at about 1.25 Hz in the 105 nm band width of a cold 878 nm filter. The LEDs are controlled by programmable power supplies providing up to 100mA of current. Leakage current is reduced to negligible levels by setting supply voltage and current to zero simultaneously to ensure that the LED is truly turned off when darkness is required. The combination of thermal regulation of the heat sink and 100 ppm/K control of current, is found to deliver 0.02% peak to peak variation over 3 hours in the mean signal (after powering on and settling) measured with a H2RG NIR detector having 1.7µm cutoff wavelength. If necessary, the LED heat sinks may be cooled by a remote fan at the end of a 4" hose extracting air the exterior to avoid disturbing the thermal environment within the projector.

The 2.5" exit port of one integrating sphere illuminates a target mask on a six-axis stage. The other sphere provides flat illumination or illuminates a second, stationary mask. A 2" non-polarizing fused silica beam-splitter cube is used to combine the beams from the spheres. Its λ/8 surface accuracy does not significantly impact optical performance, due to proximity to the focal plane and relatively slow beam. The main chromatic effect of the beam combiner is a small dependence of focus position on wavelength, about 0.3 µm per nm of Δλ in the NIR. For instance, the focal shift at the edges of our J-band filter (1250 ± 90 nm) is ~30 µm, just within our estimated 31µm depth of focus at f/8. This effect is only an issue at visible wavelengths. A greater concern is ghosting which overlays defocussed images on the primary image, thus reducing contrast and creating a faint but coherently structured background, which could affect weak lensing ellipticity estimates, for example. To mitigate ghosting, the beam combiner is coated to reflect less than 0.5% of the light from 700nm to 1100nm. The unused bottom face of the beam combiner receives 50% of the light from the flat-field sphere mounted on the top face. This light is absorbed by an Acktar Metal Velvet Black surface, which reflects less than 1% of the light that could otherwise propagate into the system by reflecting back to the target mask. The mask also has an anti-reflective coating on the chrome (5% at 700 nm, unknown in IR). Assuming 5% mask reflectivity, two passes through the 50% beam combiner, and 1% reflectivity of the Acktar coating, the flux by this path should be ~0.013% of the main beam. An internal reflection from the bottom face of the beam combiner back to the mask contributes an additional ~0.0063%. To eliminate ghost images and defocus from the beam combiner entirely, it can be removed or replaced by a mirror to select only one sphere. The beam combiner is only required for emulations that need two beams to be used concurrently (e.g. exoplanet transits, see 3.2).

Exposure times are set using a rolling shutter on CMOS detectors. For CCDs requiring a global shutter, there is an ample 13 cm of back focal distance from the turntable where a mechanical shutter may be placed. The LED and lamp controls are scriptable and provide some capability to time the stimulus, but the "off" states of these sources do not stop NIR backgrounds. Dark frames for NIR detectors are acquired using an opaque aluminum disc in one slot of the cold filter wheel.

## 2.3 Target Masks

Target masks have been procured from commercial vendors serving the semiconductor industry. e.g. HTA Photomask Inc. have supplied input masks etched in a wide range of "scenes" or patterns with 50 nm spatial resolution and 1 μm minimum feature size. The simplest of these patterns are thousands of small circular or elliptical apertures with 100% contrast, uniformly spaced across the whole image. Masks with sub-wavelength dot matrix patterns have been made by JPL's Microdevices Laboratory to produce controlled intensity profiles emulating emission and absorption features in planetary spectra (see Fig. 4). The same technique would enable emulations of galaxies with various shapes, sizes, and surface brightness profiles. The masks are etched chrome on quartz, optionally with an anti-reflective coating to reduce ghosting and scattered light. Multiple patterns are laser written on 125 mm square substrates and diced into 50mm squares. HTA claims a typical mask flatness deviation of 4 μm peak to valley, with minor changes caused by dicing and stresses from AR coatings.

"Spot" masks are made with a range of aperture sizes and spacings to adjust to the optical PSF size and to maximize the number of non-overlapping images. Apertures much smaller than the PSF are effective point sources and insensitive to etching asymmetries. For an aperture spacing of e.g. 10 pixels, 40,000 point sources can be imaged concurrently over a 2Kx2K pixel detector. *This multiplex advantage is essential since statistical averaging must be employed during the emulation to reveal low-level systematics, as is often the case when observing.* The adoption of a regular grid of aperture positions is very useful when acquiring data since many images can be aligned with rows or columns to facilitate windowed readout or to test dependencies on pixel alignment. An aperture pitch that is a non-integer number of pixels assures that a range of pixel alignments are sampled in each image. The beat frequency between aperture pitch and pixel pitch is fine tuned by selecting the mask rotation angle.

So that masks can be installed reproducibly, they are mounted in a frame with three ball to V-groove interfaces set at 120 degrees creating a kinematic interface. Frames are mounted on a 6-axis stage driven by stepper motors (Thorlabs Nanomax 604). The vendor's stepper motor resolution specification is 60 nm positioning resolution, but in practice we see 1-2 μm reproducibility as measured by the average translations of thousands of point sources. This is smaller than a typical pixel but not sufficient to dither at precise intervals, e.g. scanning in 1/2 pixel steps. When reconstructing the PSF from undersampled data, pseudo-random dithers have been used to compensate for this 1-2 μm positioning resolution. Subsequently, a piezo controlled tip-tilt stage was installed under the fold mirror, which provides ~50nm closed-loop and ~5nm open-loop positioning over 0.5mm, but it introduces a small rate dependence on field position. The 4 mm mask travel range in X, Y and Z is ample for mapping detector properties and focus control. Pitch, yaw, and roll motion up to 0.10 rad can be executed with better than 25 µrad resolution. This allows detector tilts to be matched to ~1 µm accuracy over a 40 mm field. Larger rotations are achieved using the turntable on the detector pedestal.

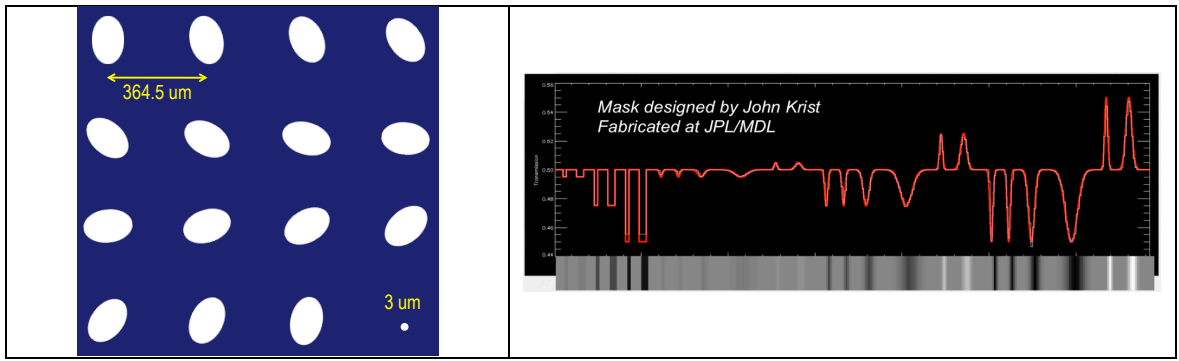

Fig. 4. Example mask templates. LEFT: Elliptical apertures of varying orientation plus a 3 µm "pinhole" to aid focus (not to scale). This cell is repeated to fill a 50x50 mm$^2$ mask. RIGHT: Strips of varying transmission and width (emulating a spectrum) are achieved by sub-wavelength dot matrix etching. A simple "toy" spectrum with a few widths and heights is shown; more realistic spectra can be likewise designed.

### 2.4 Image stabilizer

Image motion due to mechanical vibration or seeing increases the PSF width and in most cases will introduce some ellipticity. Although various emulations can compensate by adjusting exposure times or averaging over many exposures, reducing image motion increases signal to noise for measuring pixel-scale effects and expedites data acquisition. Time series of centroids from spot grids show that the mean position across the detector has an RMS displacement of about 1 µm, with most of the power coming from frequencies lower than 1 Hz. One strategy for counteracting this motion is to use the fold mirror as a tip-tilt compensator, i.e. removing the common-mode image motion by controlling the mirror angle in a closed loop with a guider camera. This has the added benefit of enabling precise, commanded image motions to emulate pointing jitter during exposures or dithering between exposures.

To create a projector guider channel, a 1" prism with aluminized hypotenuse picks off the edge of the beam after the fold mirror, just before the detector dewar window. Light is directed to a "scientific CMOS" camera (Andor Zyla) to the side of the detector pedestal. Two mounts for the pick off mirror are provided. One allows light at the edge of main field to be directed to the CMOS camera so that it shares the same target as the detector under test. The other mount places the pickoff mirror just outside the 40mm f/11 field. In this setup, a separate target with similar fold mirror at the input end of the light path provides a separately illuminated target for the CMOS camera so that the stabilizer's illumination is insensitive to the illumination used for the main beam. In this case the CMOS camera can be illuminated with a separate source (e.g. 650nm light) that falls shortward of the passband of NIR sensors under test. Most of the guider light path overlaps with the main beam so that the two channels will see most of the same seeing and vibrations; however, decorrelated motions where the light paths diverge (after the pick off or near the targets) will limit the image stabilizer performance.

The CMOS camera senses image motion at up to ~100Hz by combining centroid measurements of several hundred or even thousands of spots. The centroid calculations are performed by a Graphics Processing Unit (NVIDIA GTX 580). A piezo driven tip-tilt stage (Physik Instruments Model S-340.ASL) with built in strain gauge sensors provides closed-loop control of the fold mirror over a 2mrad range with 0.2µrad repeatability/stability. The moment of inertia of the mirror (354 g) plus cell (3D printed in lightweight carbon-reinforced Nylon-12) are kept low to avoid limiting bandwidth. The servo control loop bandwidth is then limited by the readout speed of the CMOS camera, which is set by the size of the selected subregion: 76 Hz for a $256^2$ pixel subregion or 13 Hz for $2048^2$ pixels, for example.

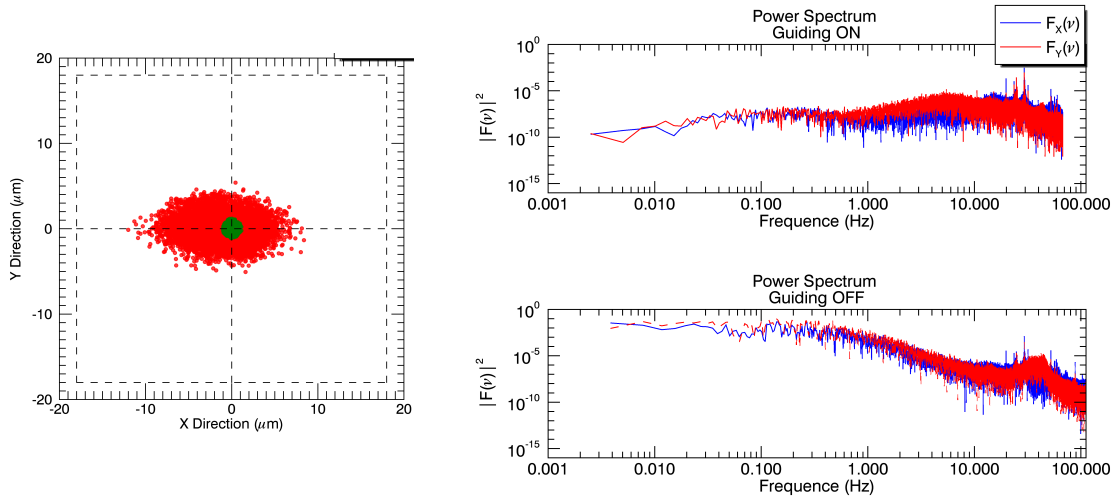

Fig. 5. LEFT: Scatter plot showing the average centroid displacement (sampled over 6 minutes) of a spot grid viewed by the guider camera with (green) and without (red) image stabilization enabled. This stabilization test achieves an order of magnitude reduction in motion and ellipticity. RIGHT: Power spectra of the centroid time series showing that most of the improvement comes from correcting low frequency motion.

Fig. 5 shows that initial tests of the image stabilizer system achieve significant reductions in image motion *as measured by the CMOS guider camera*. Not only do we reduce the RMS displacements from 1 μm to 0.1 μm, but there is also a drastic reduction in ellipticity, which improves shape measurement studies. Unfortunately, we do not yet see significant motion reduction in the science channel (detector being tested). Therefore, we are currently limited by a noise source that decorrelates between the two detectors. This conclusion is corroborated by differential measurements of image motion across the focal plane. At time of writing, our working theory is that cold air around the dewar is falling into the light paths, creating uncorrelated seeing in the two channels. Efforts to test and mitigate this effect are ongoing.

## 3  Example emulations

There are two basic types of emulation experiments. The first is a targeted investigation of a known (or hypothesized) detector effect in order to characterize or model that effect. The second type is to verify the precision of an intended science measurement (e.g. photometric stability of point sources) by mimicking that measurement. If the expected precision is not reached, that may signify a systematic error that was previously unknown or insufficiently calibrated. Emulations of both types are described below.

### 3.1  Weak Gravitational Lensing Measurement

The aim of this laboratory emulation is to show how well ellipticity caused by the image sensor can be measured and corrected. If $I(r)$ is the intensity profile of a source image, then common estimators for ellipticity are *$e_1$* and *$e_2$*, with:

$$Q_{ij} \equiv \frac{\int d^2r\, I(r)w(r)(r_i - \bar{r}_i)(r_j - \bar{r}_j)}{\int d^2r\, I(r)w(r)} \qquad \text{Equation 1}$$

$$e_1 = \frac{Q_{xx} - Q_{yy}}{Q_{xx} + Q_{yy}}; \qquad e_2 = \frac{2Q_{xy}}{Q_{xx} + Q_{yy}} \qquad \text{Equation 2}$$

where $i, j$ correspond to either axis of the pixelated image, and $\bar{r}_i$ is the weighted image centroid (1st moment). The weighting function $w(r)$ is comparable in shape to the optical PSF and ensures that the integrals converge in the presence of noise. The components $e_1$ and $e_2$ are known as the "plus" and "cross" polarizations of ellipticity; this parameterization spans the same space as an ellipticity amplitude and orientation ($e, \theta$) but avoids discontinuity at $e=0$ that prevents averaging of many sources. The goal for weak lensing surveys (e.g. WFIRST, Euclid, LSST) is to measure the correlation function $\xi(\theta)$, which is proportional to the mean product of ellipticities of all galaxy pairs separated by an angle $\theta$ on the sky. Biases in $\xi(\theta)$ must be reduced to $O(10^{-7})$ to avoid biasing dark energy parameter measurements beyond the statistical uncertainty achievable from hundreds of millions of galaxy images (which have a mean-square ellipticity of ~0.2).

Key goals for a weak lensing emulation are to show that ellipticity measurements are sufficiently stable in time and can be partitioned between optics and detector by separating the ellipticity into moving and stationary components when the detector is translated or rotated in the focal plane. One can then show how the detector contribution to the ellipticity changes as a function of spot intensity, pixel alignment, etc. The effectiveness of detector calibration algorithms in improving the quality of the ellipticity measurement can be investigated, as can the effect of choice of plate scale, observing cadence and dither strategy. Instrumental sources of ellipticity for WFIRST will be measured using images of stars as calibrators. Systematic differences in PSF measurements for stars and for galaxies (fainter, lower contrast) are thus a key area of investigation through emulation.

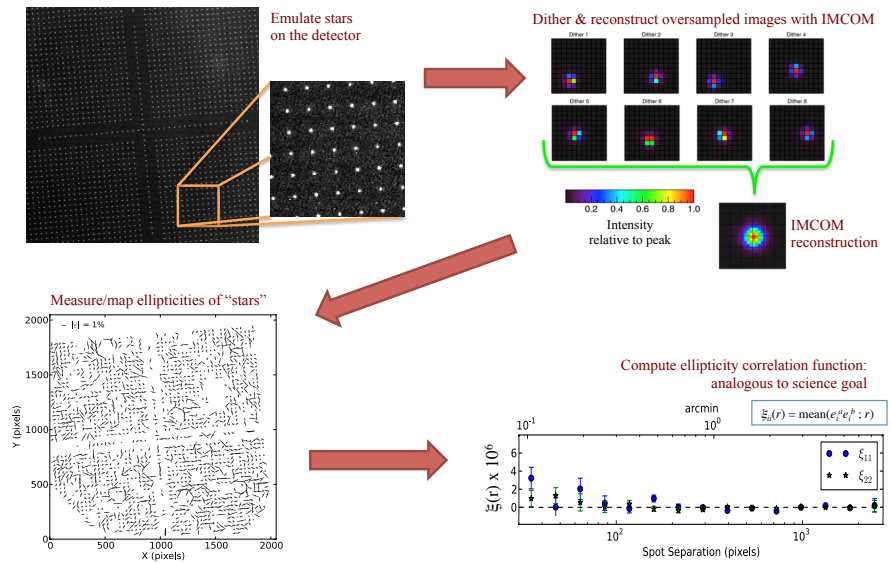

Fig. 6. Flow diagram of a weak gravitational lensing emulation. A grid of "stars" is projected onto an H2RG detector at a wavelength and f-number that mimics an undersampled space mission. Fully-sampled images are reconstructed from multiple dithered images so that ellipticity can be measured and mapped. The ellipticity correlation functions are computed and compared to weak lensing error tolerances [21].

In order to maximize their field of view, space missions and concepts such as Euclid and WFIRST are designed to be undersampled. Shapes cannot be reliably measured from single, undersampled expo-

sures – a galaxy or PSF profile must be reconstructed from dithered images allowing the pixel centers to sample different locations. PPL emulates this strategy by translating the image mask between exposures and using the IMCOM algorithm developed for WFIRST [26] to reconstruct an oversampled image for which we can measure an ellipticity [21].

Even though the Offner design has very low aberrations, the dominant one is astigmatism, for which ellipticity is proportional to defocus. Mask tilt can be adjusted to focus the largest possible region of the image, but the detector and mask surfaces are not perfectly flat. For the detector, we typically measure < 20 µm peak to valley with a Zygo interferometer. To remove ellipticity due to defocus, the best focus image for each spot can be selected from a through-focus scan. The ellipticity map obtained from the shape of each spot at its best focus position is what would be expected from a flat detector and mask with no tilt. With thousands of spots imaged at multiple focus positions and dither positions, data volume can grow quickly. Our analysis pipeline benefits from parallelizing the data reduction on a 144-core computing cluster

### 3.2 Transit Spectrophotometry Experiments

JWST (and possibly FINESSE) will detect the atmospheres of exoplanets by continuously observing low-resolution, NIR spectra as the planet transits the host star. The high signal to noise ratio (better than 50 parts per million) required to separate the absorption and emission features of the planetary spectrum from the continuum spectrum of the host star will be achieved by co-adding many spectra. Principal component analysis (PCA) will be used to measure the correlations between the spectroscopic signal and system disturbances such as pointing jitter. While PCA eliminates the need for conventional calibration, it can only deliver photon-shot-noise limited performance if the sources of the errors are correctly identified and measured with sufficient precision to make the de-correlation possible.

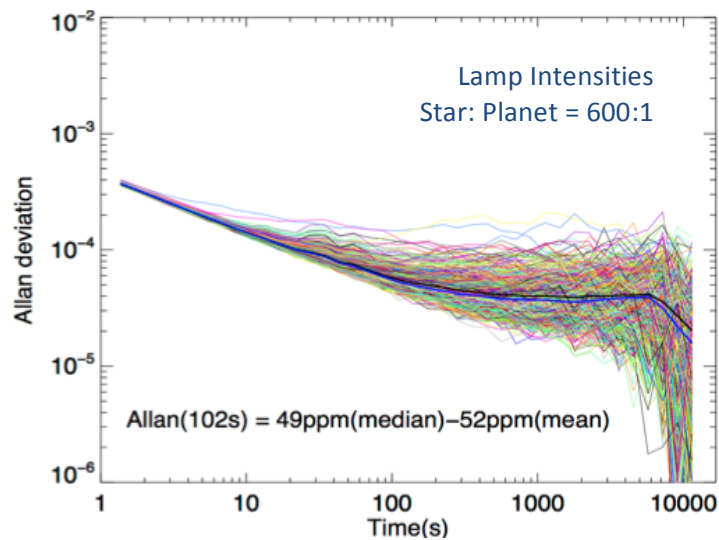

Fig. 7. Example PPL demonstration by D. Touli *et al* showing photometric stability of an H2RG detector observing an emulated exoplanet spectrum (as in Fig. 4) combined with a bright background [22,23]. Thin curves represent Allan variances computed from single pixel columns. The mean (median) is shown as a thicker black (blue) curve. Decorrelation improves the mean to ~20ppm, which is sufficient to detect an Earth-like exoplanet transit.

Emulation of spectrophotometry uses a beam combiner to optically superimpose images representing the stellar and planet spectra. Separate stabilized lamps for each source are adjusted to have the desired relative intensity. To date we have emulated only the intensity distribution of the expected spectrum and not the wavelength variation. Flat illumination emulates the continuum spectrum of the star and a grey scale mask emulates the images of emission and absorption lines with a range of widths and depths. The planetary transit signal is created by either moving the star mask out of the pixel region representing the combined spectrum or by modulating the lamp intensity. Pointing jitter can be created with the piezoelectrically controlled tip-tilt mirror. In the future, a dispersive grating could be fabricated to replace the secondary mirror, allowing us to emulate true multi-chromatic images.

Many hours of short exposures having relatively high intensity are acquired continuously for periods in which there is nominally no disturbance. Allan variance (a measure of stability over a given time interval) is computed to show the signal coaddition period at which the variance exceeds the photon shot noise limit. The extension of shot noise limited Allan Variance to longer coaddition depth and thus higher statistical precision is demonstrated as correlations with various kinds of disturbances are measured by PCA and removed. These disturbances may be irreducible features of the test system, such as fluctuations in lamp brightness and ambient temperature, or induced disturbances such as pointing (image motion), focus changes, detector temperature, or detector bias voltages. The emulation allows the sensitivity to various system parameters to be measured so that the source of the noise floor can be identified.

### 3.3 Rapid Intra-pixel Response Characterization

Intra-pixel response variation is an example of a detector effect that cannot be calibrated by conventional analysis of dark and flat images. Flat-field calibrations only normalize the mean Quantum Efficiency (QE) of each pixel to a common level. When QE varies within a pixel, measurements of high-contrast images (stars, galaxies, spectral lines) will depend on their location relative to the pixel grid, particularly if the optical PSF is undersampled.

One possible source of intra-pixel response variations is an infamous "feature" of HgCdTe arrays sometimes called the "crosshatch" pattern. The crosshatch manifests in flat field images as a fixed pattern of high frequency QE variations along 3 directions. The bands in Fig. 8 have wave-vectors pointing along 0, 45, and 112 degrees. The axes are generally thought to be related to the crystal structure of HgCdTe. The amplitude of the crosshatch varies spatially, and it may vary among detectors of a single device lot; thus, projects may consider it when selecting candidate devices for an instrument. The pattern is evident in WFC3/IR, WISE, JWST, and in candidate detectors for Euclid and WFIRST. It may induce correlated errors in photometry, astrometry, and shape measurements if not properly calibrated.

The Euclid Near Infrared Spectrophotometer (NISP) has a goal of ~1% relative photometry (RMS), and only a portion of the error budget may be allocated to the H2RG detectors. With f/10 optics and 18 µm pixels, it is strongly undersampled at wavelengths below 1.8 µm, then weakly undersampled up to its 2.3 µm cutoff, and thus sensitive to intra-pixel variations. In recognition of the risk to meeting science requirements, an engineering grade H2RG was lent to PPL in order to investigate the impact of the crosshatch pattern on NISP photometry. A basic emulation strategy is to quantify spatial variations in photometry in post-calibrated (flat-fielded) images.

A grid of ~18,000 point sources (15.25 pixel pitch) was projected with a Euclid-like PSF (f/11, 1 µm illumination). The grid was scanned over 3 pixels in 6 µm steps (1/3 pixel), corresponding to the band limit of the PSF under these conditions, i.e. measurements are insensitive to QE variations at higher frequencies. Aperture photometry in a 3-pixel radius was measured for each spot after applying standard calibrations: dark and background subtraction, flat-field correction, nonlinearity correction (N.B. aperture photometry is insensitive to IPC). Spots near known bad pixels were excluded. The relative scatter in each spot's photometry was computed, and from this we subtract each spot's baseline scatter from a series of images where the grid was kept at a fixed position.

Differences in photometric variation across the detector (Figure 6) show a clear demarcation between the strong crosshatch and weak crosshatch regions. Typical spots in the strong crosshatch region pick up 1-2% additional scatter in photometry when they sample different locations. In the weak crosshatch region, there is no additional scatter on average, but the effects may be detectable in isolated regions. This result implies that the crosshatch arises from response variations that are not accounted for by flat-field calibration. Thus Euclid can mitigate photometric errors by selecting detectors with weaker crosshatch. Characterization experiments with a single, highly focused beam can obtain higher resolution maps of the sub-pixel structure [18,19]; however, by approximating the NISP PSF, our measurements are more directly relevant to the Euclid requirements. Nevertheless, using extended scans, we can infer a map of QE across the entire detector down to a scale set by the PSF bandlimit (6 μm or 1/3 pixel in this example). The multiplex advantage of projecting thousands of sources has thus allowed PPL to provide a rapid and comprehensive look at how sub-pixel structure affects photometry. We will present a comprehensive report of this study in an upcoming paper.

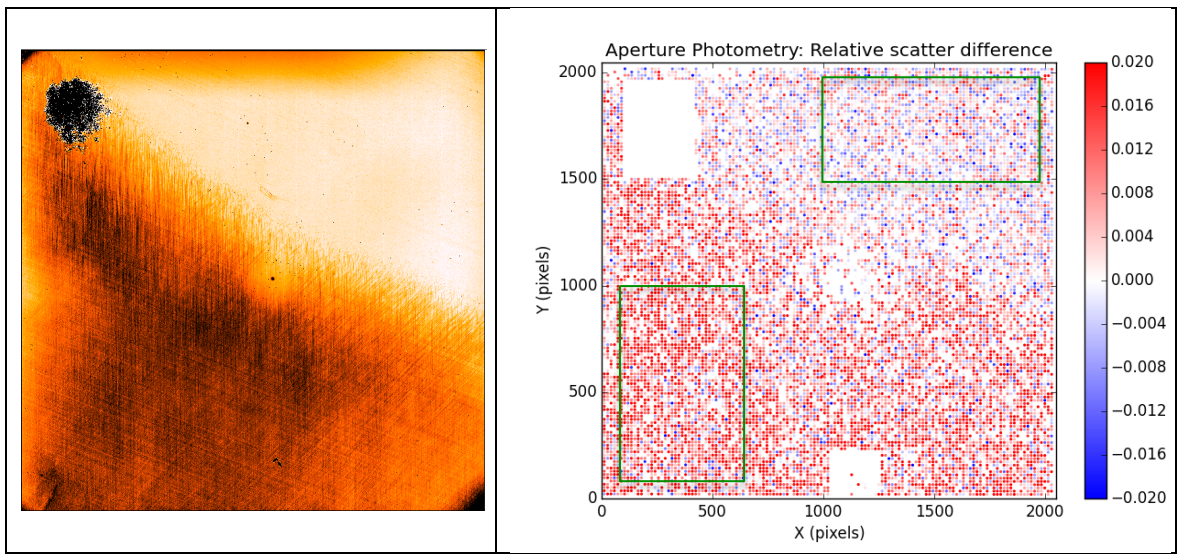

Fig. 8. LEFT: Stretched flat-field image from an engineering grade H2RG exhibiting a strong crosshatch pattern on one side. RIGHT: Map of photometric variations of Euclid-like point sources projected on the same detector. Each dot represents the change in the relative scatter (stddev/mean) of aperture photometry for a spot image when it is scanned in a small region (3 px in 1/3 px steps) relative to when it is stationary. Averaged over the regions marked by green boxes, we find a 1% increase in scatter in the strong crosshatch region and no increase in the weak crosshatch region. N.B. the averaging suppresses isolated features in the weak region that are visible in the flat.

### 3.4 Brighter-Fatter Effect measurement

The "brighter-fatter" (BF) effect is a phenomenon wherein the PSF appears to widen with increasing fluence. It is a pernicious effect for weak gravitational lensing analyses, which use the PSF measured from bright stars to account for optical shape distortions imparted to faint galaxies. BF was first discovered in CCDs by the Dark Energy Survey, which needs sub-percent precision in PSF size measurements. The leading explanation for BF in CCDs is that electric fields from charges accumulating in the bright center of a PSF image are deflecting subsequent charges away, thereby shifting the effective pixel boundaries as charge builds up during an exposure. PPL has provided evidence for a BF effect in an H2RG detector, the existence of which is less established [14]. A pixel-shifting mechanism for CMOS detectors is plausible since each pixel is associated with a PN-junction whose depletion region physically shrinks as it collects charge.

Whereas in CCDs, BF is identified by comparing multiple PSF images with varying exposure levels, we can use the non-destructive read feature of the H2RG to look for changes in individual images as we "sample up the ramp". In the absence of the BF effect, multiple reads in an exposure should show that the flux in each pixel is constant once nonlinearity has been calibrated. A signature of the BF effect in a CMOS detector would be a detection of flux "transfer" from a bright central pixel of a PSF to the fainter neighboring pixels. This occurs non-linearly in time, unlike the effect of IPC, for which a fixed fraction of signal in a pixel is shared among neighboring pixels.

Using the same projector setup as in 3.3 (18,000 spots, f/11, 1 μm illumination), we first use flats to calibrate the nonlinearity in each pixel. The BF effect would be suppressed in flat images since they contain little contrast between pixels. We then project the spot grid and identify spots with centroids within 0.1 pixels of a pixel center. In these cases, the flux is strongly concentrated in the central pixel – the optical full-width-half-maximum (FWHM) is 11 μm, increased to ~14.5 μm by lateral charge diffusion and seeing (see Fig. 9). The ramps of these spots are averaged over 90 exposures (preceded by 10 exposures to allow charge trapping effects to equilibrate. We then infer pixel fluxes versus time from the differences between successive reads (excluding the reset frame) and average over all spots.

Fig. 9 shows that we detect an apparent BF signature: over the course of the exposures, flux in the central pixels decreases while flux in the 8 nearest neighbor pixels increases. The changes are not quite equal and opposite, as would be expected from charge conservation. Confounding factors may include inadequate calibration of nonlinearity, including possible nonlinear IPC. Assuming we have inferred the correct order of magnitude – a percent level redistribution of charge – a BF effect of this size must be accounted for by e.g. WFIRST, which can only tolerate $O(10^{-3})$ errors in PSF size. Additional details of this work will be described in an upcoming paper.

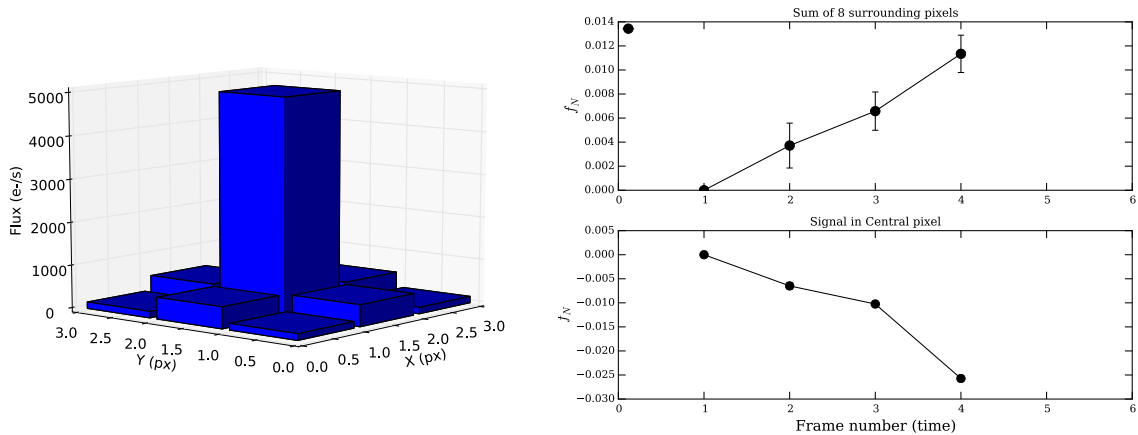

Fig. 9. LEFT: Flux profile of a representative spot (f/11, 1μm light, 18 μm pixels) showing nearly 10:1 contrast between the center and nearest neighbor pixels. RIGHT: Flux vs. frame (frame time = 3s) measured from successive non-destructive reads "up the ramp", averaged over ~700 spots and 90 ramps per spot. $f_N$ is the change in a pixel's flux relative to the start of the exposure, divided by the time-averaged total flux of the spot. With pixel-wise non-linearity corrections applied, the central pixels appear to lose flux as the image integrates while the neighbor pixels see a comparable flux increase, as expected for the brighter-fatter effect.

## 4    Conclusions

The PPL's testbed is fully functional and capable of emulating a wide range of observing scenarios for space and ground telescopes with many optical characteristics that exceed the performance of planned missions. Ongoing PPL experiments are informing detector selection and calibration strategies for Euclid and WFIRST, which have challenging photometric and shape measurement requirements. Past experiments have provided proof of concept studies for exoplanet transit detection with JWST and FINESSE. In addition, PPL has enabled demonstrations of guider cameras for Keck, PPL's own stabilization system, the Wafer-Scale imager for Prime (WaSP) guider for the 200" Hale Telescope at Mt. Palomar, and soon the SuperBIT balloon-borne telescope. PPL can rapidly characterize detectors over large areas (40x40 mm$^2$), allowing specific detector effects to be investigated or calibration procedures to be validated by assessing impact on scientific measurements such as photometry, spectroscopy, astrometry, and shape measurement. The versatility of the projector allows new experiments to be designed and executed quickly to support studies for ongoing missions or future proposals.

Work is ongoing to further increase the projector stability and contrast of scenes. These improvements will allow us to more closely emulate space-based observations, and they will increase signal to noise for a variety of measurements, which accelerates data acquisition. A priority for future work is identifying and mitigating sources of image motion. Initial tests of the image stabilizer system have shown we can in principle reduce common mode motions to 0.1μm or better; however, the limiting source is currently not correlated over the wide field of view. Image contrast can be improved by experimenting with baffles to reduce scattered light and different materials to attenuate ghost images from secondary reflections. Expanding the range of observations we can mimic is also of interest, and we strive to stay informed about the challenging detector needs of the astronomical community in order to support present and future missions.

## Acknowledgements

We sincerely thank members of Caltech Optical Observatories who contributed to the construction and maintenance of the PPL testbed: Dave Hale, Jennifer Milburn, Hector Rodriguez, Patrick Murphy, Michael Feeney, Justin Belicki, Alex Delacroix, and former team members Eric Jullo and Viswa Velur, who were key players in the early stages of development. Thanks to Warren Holmes and the Euclid detector working group for lending the H2RG detector used in sub-pixel and brighter-fatter effect measurements. This work was carried out at the Jet Propulsion Laboratory, California Institute of Technology, under a contract with the National Aeronautics and Space Administration.